\UseRawInputEncoding
\documentclass[pdftex,twocolumn,epjc3]{svjour3}          % twocolumn

\RequirePackage[T1]{fontenc}

\smartqed  % flush right qed marks, e.g. at end of proof

\RequirePackage{graphicx}
\RequirePackage{mathptmx}      % use Times fonts if available on your TeX system
\RequirePackage{placeins}
\RequirePackage{flushend}
\RequirePackage[numbers,sort&compress]{natbib}
\RequirePackage[colorlinks,citecolor=blue,urlcolor=blue,linkcolor=blue]{hyperref}

\journalname{Eur. Phys. J. A}

\begin{document}

%\title{Heavy flavour suppression with detailed balance at RHIC and LHC energies}
\title{Effect of thermal gluon absorption and medium fluctuations on heavy flavour nuclear modification factor at RHIC and LHC energies}


\author{Ashik Ikbal Sheikh\thanksref{e1,addr1}}

%\thankstext[$\star$]{t1}{Thanks to the title}
\thankstext{e1}{e-mail: ashikhep@gmail.com/asheikh2@kent.edu}

\institute{Department of Physics, Kent State University, Kent, OH 44242, USA\label{addr1}
}

%\institute{First Address, Street, City, Country\label{addr1}
%	\and
%	Second Address, Street, City, Country\label{addr2}
%	\and
%	\emph{Present Address:} Street, City, Country\label{addr3}
%}


\maketitle

\begin{abstract}
 The presence of thermal gluons reduces the stimulated gluon emission off a heavy quark propagating in the quark-gluon plasma (QGP) while absorption causes reduction of the radiative energy loss. On the other hand, the chromo-electromagnetic field fluctuations present in the QGP lead to collisional energy gain of the heavy quark. The net effect of the thermal gluon absorption and field fluctuations is a reduction of the total energy loss of the heavy quark, prominent at the lower momenta. We consider both kind of the energy gains along with the usual losses, and compute the nuclear modification factor ($R_{AA}$) of heavy mesons, viz., $D$ and $B$ mesons. The calculations have been compared with the experimental measurements in Au-Au collisions at $\sqrt{s_{NN}} = 200$ GeV from STAR and PHENIX experiments at the RHIC and Pb-Pb collisions at $\sqrt{s_{NN}} = 2.76$ TeV and $5.02$ TeV from CMS and ALICE experiments at the LHC. We find a significant effect of the total energy gain due to thermal gluon absorption and field fluctuations on heavy flavour suppression, especially at the lower transverse momenta.
\end{abstract}

\section{Introduction}
\label{intro}

 The heavy-ion programs at the BNL Relativistic Heavy Ion Collider (RHIC) and the CERN Large Hadron Collider (LHC) are aimed to create and characterize the quark-gluon plasma (QGP), a deconfined state of QCD matter. Heavy quarks (charm and bottom) are mostly produced by the primordial hard scattering processes in the early stages of the collisions. They may be produced in the QGP, if initial temperature of QGP is high enough than the mass of the heavy quarks. The gluon splitting process can also produce the heavy quarks at the late stage~\cite{Xing:2019xae}. However, the total number of heavy quarks mainly becomes frozen in the early time of the collision history. The energy dissipation of heavy quarks inside the QGP medium is considered as one of the promising probes for the characterization of QGP and it remains an ever active and important field of research~\cite{Xing:2019xae,TG, BT, Alex, MG,mgm05,dokshit01,dead,Fochler:2010wn,alam10,abir12,Abir10,das10,Fochler:2008ts,Gossiaux10,horowitz10,wicks07,GLV,Rapp,Dong:2019unq,Santosh14,Li18,LiWang,Hees2,Gossiaux,Xu18,WKe,PP,Ours1,Ours3,jamil10,armesto1,armesto2,B.Z,W.C,Vitev,yd,skd,Uphoff11a,Uphoff11b,Fl,AJMS,KPV,KPV1}. The energy loss of energetic heavy quarks while propagating through the QGP medium is manifested in the high transverse momentum, $p_T$, suppression of heavy flavour hadrons ($D$ and $B$ mesons). Observables characterizing the depletion of heavy flavour hadrons produced in nucleus-nucleus ($A$-$A$) collisions with respect to those produced in $p$-$p$ collisions are measured at the RHIC~\cite{Rhic1,Rhic2} and the LHC~\cite{ALICE_DNpart,ALICE_D,CMS_D_2TeV,CMS_D_5TeV,CMS_B_5TeV}. The energy loss mechanisms are governed via collisions and bremsstrahlung gluon radiations due to the interaction of heavy quarks with the medium partons. The collisional~\cite{TG,BT,Alex,PP} as well as  radiative~\cite{MG,mgm05,dokshit01,dead,wicks07,GLV,armesto2,B.Z,W.C,Vitev,AJMS,KPV} energy loss have been reported extensively in the literature. 
 
 Generally, gluon radiation induced by multiple scattering of an energetic parton propagating in a dense medium leads to medium induced parton energy loss. The theoretical studies of heavy quark radiative energy loss
 via such mechanisms can be found in Refs.~\cite{MG,mgm05,dokshit01,dead,wicks07,GLV,armesto2,B.Z,W.C,Vitev,AJMS,KPV}. The calculations of the radiative energy loss of heavy quarks incorporate many interesting properties due to the non-Abelian Landau-Pomeranchuk-Migdal (LPM) interference effect and suppression of soft gluon emission from heavy quarks due to the heavier masses which is known as the dead cone effect~\cite{dokshit01,dead}. Because of the presence of thermal gluons in the hot QGP medium, along with the stimulated gluon emission, the propagating parton absorbs thermal gluons as well. This thermal absorption causes a reduction of the radiative energy loss as argued in Ref.~\cite{wangwang}. On the other hand, QGP being a statistical system of mobile coloured partons, stochastic fluctuations of the produced chromo-electromagnetic field are expected in the system. Such microscopic fluctuations affect the response of the medium on the influence of external perturbations. When heavy quarks propagate, these fluctuations cause fluctuations in the velocities of the heavy quarks which result in collisional energy gain of the propagating heavy quarks. The energy gain is significant at lower momenta and it has been calculated in detail at Ref.~\cite{Fl}. These fluctuations have an important effect on heavy quark propagation in the QGP and the transport coefficient of the medium as shown in our previous work~\cite{Ours2}. In our earlier work, the effect of the thermal gluon absorption by heavy quarks was completely ignored.
 
 The net effect of the thermal absorption and chromo-electromagnetic field fluctuations is a reduction of radiative and collisional energy loss of heavy quarks, respectively. Both the effects cause net reduction in the total energy loss of the heavy quarks. This detailed account of energy loss and gain is crucial for studying the heavy flavour suppression in a hot medium. In this article, we consider the energy gains (thermal+fluctuations) and losses (collisional+radiative) by heavy quarks while computing the nuclear modification factor ($R_{AA}$) of $D$ and $B$ mesons at the top RHIC and the LHC energies. We compare our results with the experimental measurements in Au-Au collisions at $\sqrt{s_{NN}} = 200$ GeV by STAR and PHENIX experiments at the RHIC and Pb-Pb collisions at $\sqrt{s_{NN}} = 2.76$ and $5.02$ TeV by CMS and ALICE experiments at the LHC.

 The present article is organized as follows: In the next section, we briefly discuss formalism and setup which contains initial spectra and fragmentation of heavy quarks, the model for QGP evolution, initial conditions, and the calculations for heavy quark energy losses and gains. We consider here: (a) the radiative energy loss obtained in generalized dead cone approach~\cite{AJMS} (b) the collisional energy loss  by Thoma and Gyulassy (TG) formalism~\cite{TG} (c) the energy gain due to thermal gluon absorption~\cite{wangwang} and (d) the chromo-electromagnetic field fluctuations as prescribed in Ref.~\cite{Fl}. In Sec.\ref{sec3}, we discuss the results. Section~\ref{sec4} is devoted to summary and conclusion.

\section{Methodology and Setup}
\label{sec2}
\subsection{Heavy quark spectra and fragmentation}
\label{sec2.1}
 The production cross sections of heavy quarks in $p$-$p$ collisions are obtained using factorization approach with elementary cross sections calculated up-to leading order (LO) and next-to-leading order (NLO) CT10 parton density functions \cite{CT10}. For heavy-ion collisions, i.e., Au-Au and Pb-Pb, the shadowing effect is taken into account by using the {spatially dependent} EPS09~\cite{EPS09} parton distribution function sets. After inclusion of nuclear shadowing effect for a given centrality class corresponding to a range of impact parameters $b_{1}$ and $b_{2}$, the differential cross section is calculated. The spectrum in Au-Au and Pb-Pb collisions are obtained by shifting the calculated differential cross section by the momentum loss $\Delta p_{T}$. Peterson fragmentation function~\cite{Peterson} is used for the fragmentation of $c$ quarks into $D$-mesons and $b$ quarks into $B$-mesons,  with parameters $\epsilon_{c} = 0.016$ for $c$ quarks and $\epsilon_{b} = 0.0012$ for $b$ quarks. For further details of the production and fragmentation of heavy quarks we refer the readers to Ref.~\cite{KPV1}.

Finally the nuclear modification factor $R_{AA}$ is computed as:
\begin{equation}
 R_{AA}(p_{T},b_{1},b_{2}) = \frac{\Big(\frac{d^{2}\sigma(p_{T},b_{1},b_{2})}{dp_{T}^{2}dy}\Big)_{A-A}}{\int_{b1}^{b2}d^{2}bT_{AA}\Big(\frac{d^{2}\sigma(p_{T})}{dp_{T}^{2}dy}\Big)_{p-p}}.
\end{equation}
where, $b_{1}$ and $b_{2}$ are the impact parameters corresponding to a given centrality class, $\frac{d^{2}\sigma}{dp_{T}^{2}dy}$ is the differential production cross-section and $T_{AA}$ is the nuclear overlap function.

\subsection{ Model for QGP evolution and initial condition}

We assume a heavy quark pair is produced at a point ($r$,$\phi$) in a heavy-ion collision, where $\phi$ is the angle with respect to radial position $\hat{r}$ in the transverse plane. After production, the heavy quark propagates through the QGP medium and lose energy during its' passage. In order to estimate the energy lost by the heavy quark, it is important to calculate the total path length traversed. If $R$ is the radius of the colliding nuclei, the path length $L$ traversed by the heavy quark inside the QGP medium is calculated as~\cite{Muller}:
\begin{equation}
L(r,\phi) = \sqrt{R^{2}-r^{2}\sin^{2}{\phi}} - r\cos{\phi}.
\end{equation}
The average distance $L(r,\phi)$ reads as,
\begin{eqnarray}
\label{eq2}
\langle L \rangle &= \frac{\int\limits_{0}^{R}rdr\int\limits_{0}^{2\pi}L(r,\phi)T_{AA}(r,b)d\phi}{\int\limits_{0}^{R}rdr\int\limits_{0}^{2\pi}T_{AA}(r,b)d\phi} , 
\end{eqnarray}
where $T_{AA}(r,b)$ is the nuclear overlap function at a given impact parameter $b$.
The effective path length of the heavy quark in the QGP of lifetime $\tau_f$ is estimated as,
\begin{eqnarray}
L_{\mbox{eff}} &= \mbox{min}[\langle L \rangle, \frac{p_T}{m_T} \times \tau_f].
\end{eqnarray}
where $p_T/m_T$ is the velocity of the heavy quark with $p_T$ ($m_T$) is the transverse momentum (transverse mass). The $L_{\mbox{eff}}$ is used in this work to estimate the heavy quark energy loss and gain.

For the QGP medium evolution in each centrality bin, we consider an isentropic cylindrical expansion with volume element $V(\tau)$~\cite{EOS}:
\begin{eqnarray}
V(\tau) &= \tau\pi(R_{tr}+\frac{1}{2}a_{T}\tau^2)^2,
\end{eqnarray}
where, $R_{tr}=R_{tr}(N_{part})$ being the initial transverse size, $N_{part}$ is the number of participant for a given centrality class, $a_T=0.1$ $c^{2}fm^{-1}$ is the transverse acceleration~\cite{Zhao} and $\tau$ being the proper time. The temperature is estimated as a function of proper time by using Lattice QCD and hadronic resonance equation of states~\cite{Huo}, and the entropy conservation condition, $s(T) V(\tau) = s(T_0) V (\tau_0)$, where the initial entropy density, $s(T_0)$, can be calculated by experimentally measured $dN/d\eta$($N_{part}$)~\cite{dndeta_alice,dndeta_cms}.
The initial volume is estimated by, $V(\tau_0) = \pi [R_{tr}(N_{part})]^2\tau_0$ with $R_{tr}(N_{part})$ being the transverse size. The $R_{tr}(N_{part})$ can be calculated by assuming the transverse overlap area, $\pi R_{tr}^{2}$, of the collision zone is proportional to the $N_{part}$. For perfect overlap, the transverse overlap area ($=\pi R^2$, with $R$ is the radius of the nucleus) is maximum and $N_{part}(=2A$, where $A$ is the mass number of the colliding nucleus) is also maximum. This implies: $R_{tr}(N_{part}) = R\sqrt{N_{part}/2A}$. For the details of the medium evolution, we refer the readers to Ref~\cite{EOS}. The medium evolution starts at initial time (initial temperature) $\tau_0$ ($T_0$) and it stops when the medium reaches freeze-out time (temperature) $\tau_f$ ($T_f$). This approach for medium evolution is followed earlier in literature~\cite{KPV,KPV1}, and we use the same set of parameters as calculated in Refs.~\cite{KPV,KPV1}. Different parameters (such as impact parameter $\langle b \rangle$, $N_{part}$, $\langle L \rangle$, $\tau_0$, $\tau_f$ and $T_0$) used in our calculations for different centrality classes are given in Table~\ref{table:1}. In this model, after having these parameters, the energy loss and gain of heavy quarks are estimated as a function of proper time, then averaged over the entire evolution of the QGP medium.

\begin{table}[h!]
\centering
\begin{tabular}{| c| c| c| c| c|}
 \hline
 $\sqrt{s_{NN}}$ (TeV) & 0.2 & 2.76 & 2.76 & 5.02  \\
\hline
 Centrality class ($\%$) &  0-10 &  0-10  & 0-100 & 0-100  \\
\hline
$\langle b \rangle$ (fm) &  3.26 &  3.44  &  9.68   &   9.65  \\
\hline
$N_{part}$  & 329 & 356  &  113  &  114 \\
\hline
$\langle L \rangle$ (fm)  &  5.63 & 5.73 &  4.16 &  4.18  \\
\hline
$T_{0}$ (GeV) & 0.303 & 0.467  & 0.436 & 0.469 \\
\hline
$\tau_{0}$ (fm/c) & 0.6 & 0.3 & 0.3 & 0.3 \\
\hline
$\tau_{f}$ (fm/c) & 3 & 6 & 6 & 6 \\
\hline
\end{tabular}
\caption{ Parameters for evolution of the QGP medium.}
\label{table:1}
\end{table}

\subsection{Heavy quark energy gain}

\subsubsection{Thermal gluon absorption }

Gluon radiation from an energetic parton propagating in a dense medium is the dominant and essential mechanism of parton energy loss. Many authors~\cite{MG,mgm05,dokshit01,dead,wicks07,GLV,armesto2,B.Z,W.C,Vitev,AJMS,KPV} calculated the radiative energy loss with various ingredients and kinematical conditions. The soft gluon emission from heavy quarks was found to be suppressed in comparison to the case of light quarks due to the mass effect, also known as dead cone effect\cite{dokshit01,dead}. The study of medium induced radiative energy loss due to the dead cone effect was initially limited only to the forward direction. Later on, improved calculations were made in Ref.~\cite{abir12} by relaxing some of the constraints imposed in the work of Refs.~\cite{dokshit01,dead}
(e.g., the gluon emission angle and the scaled mass of the heavy quark with its energy). The improved calculations~\cite{abir12} led to a compact expression for the gluon emission probability from a heavy quark and introduced a generalised version of the dead cone effect. The heavy quark radiative energy loss is computed based on the generalised dead cone approach and the gluon emission probability in Ref.~\cite{AJMS}. 

These radiative energy loss calculations consider only the medium induced gluon emission from the propagating heavy quarks. However, QGP being a hot statistical system of quarks and gluons, the temperature ($T$) of the medium produces thermal gluons. The number of thermal gluons, $N_g(|{k}|)$ grows with $T$~\cite{wangwang}: $N_g(|{k}|)=\frac{1}{e^{|{k}|/T} -1}$, where ${k}$ is momentum of the thermal gluon.
When an energetic heavy quark propagates through such a hot medium, the heavy quark absorbs thermal gluons along with emitting medium induced gluons. Naively, the energy scale associated with the stimulated emission and thermal absorption is around $T$. The process of thermal absorption causes energy gain, mostly at the lower momenta, which reduces the radiative energy loss. A quantitative estimate of the effect of the thermal absorption was made in Ref~\cite{wangwang}. The energy gain via thermal absorption with ($\Delta E^{(1)}_{abs}$) and without ($\Delta E^{(0)}_{abs}$) re-scattering by the medium partons reads as~\cite{wangwang}:

\begin{equation}
\frac{\Delta E^{(1)}_{abs}}{E} \approx \frac{\pi\alpha_{s}C_F}{3}\frac{LT^2}{\lambda_{g}E^2} \left(log\frac{\mu^2L}{T}-1+\gamma_{E}-\frac{6\zeta'(2)}{\pi^2}\right)
\end{equation}

\begin{equation}
\frac{\Delta E^{(0)}_{abs}}{E} \approx \frac{\pi\alpha_{s}C_F}{3}\frac{T^2}{E^2} \left(log\frac{4ET}{\mu^2}+2-\gamma_{E}+\frac{6\zeta'(2)}{\pi^2}\right)
\end{equation}
where $\gamma_{E} \approx 0.5772$, $\zeta'(2)\approx −0.9376$, $\alpha_s = 0.3$ is the strong coupling constant, $C_F$ is the Casimir factor, $E=\sqrt{p^2+m^2}$ is the energy of the heavy quark with $p$ and $m$ are momentum and mass of the heavy quark respectively, $L$ is the distance traversed by heavy quark, $\lambda_q$ is the mean free path, $\mu^{2} = 4\pi\alpha_{s}T^{2}\left(1+{n_f}/{6}\right)$ is the square of Debye screening mass, $n_f =2$, is the number of active quark flavours.

\begin{figure}[htb!]
	\centering
		\includegraphics[width=0.45\textwidth]{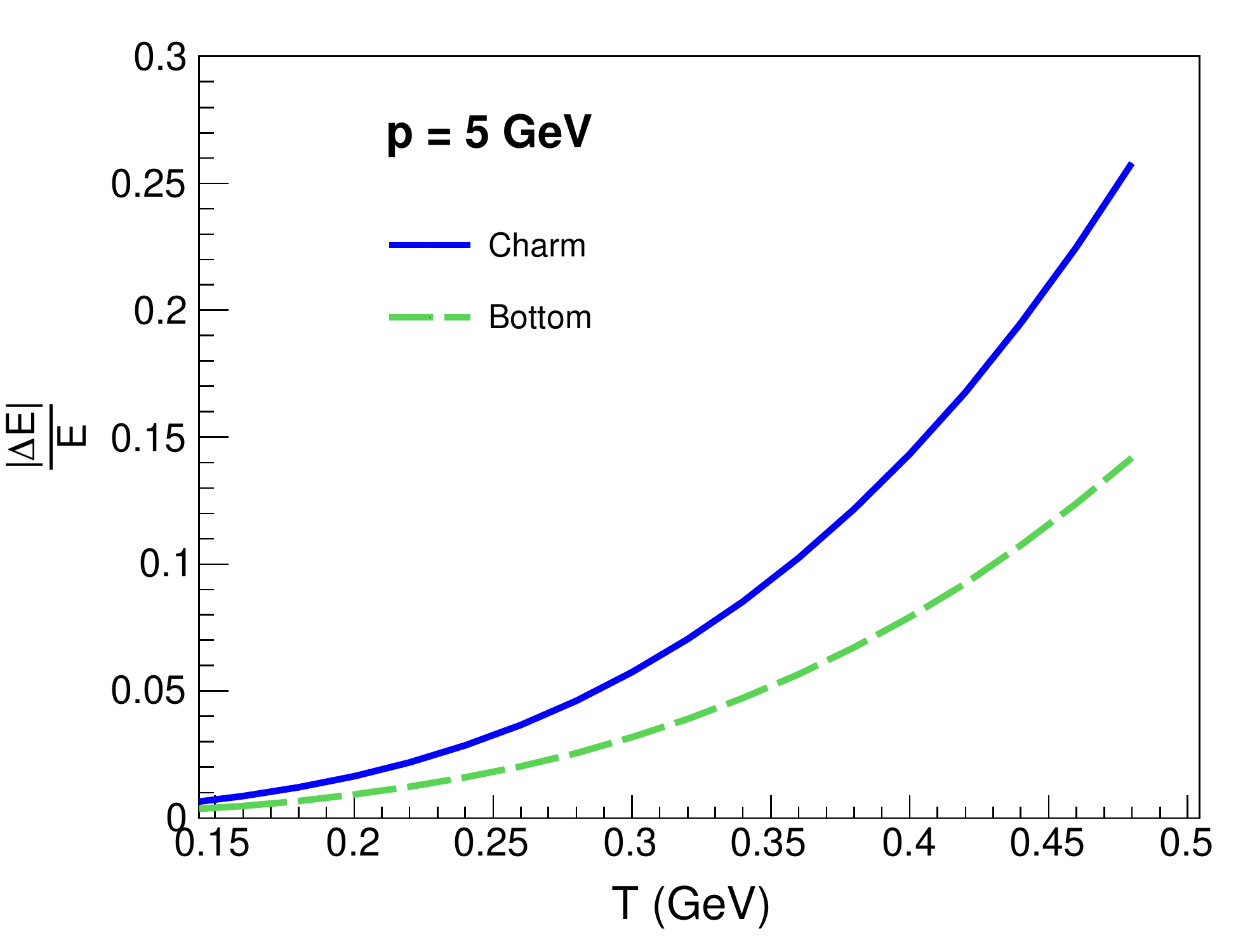}
		\caption{Fractional energy gain of a charm and a bottom quark of momentum $p=5$ GeV as a function of temperature $T$ of the QGP medium. The energy gain is due to thermal gluon absorption~\cite{wangwang}.}
		\label{delEbyEVsT_cb}
\end{figure}

The fractional energy gain due to thermal gluon absorption~\cite{wangwang} for a charm and a bottom quark of momentum $p=5$ GeV as a function of temperature $T$ is shown in Fig.~\ref{delEbyEVsT_cb}. The parameters chosen are:  $\alpha_s = 0.3$, charm quark mass $m_c = 1.25$ GeV, and bottom quark mass $m_b = 4.2$ GeV. It is observed that the thermal energy gain, both for charm and bottom quark, increases with $T$. This is because of the fact that the number of thermal gluons in the medium grows with $T$, and hence, the probability of thermal gluon absorption by the propagating charm and bottom quark increases with $T$. The bottom quark being heavier than the charm quark, this thermal energy gain is less for the bottom quark than the charm quark.

In this work, we use this radiative energy gain caused by the thermal absorption~\cite{wangwang}, and the generalised dead cone based radiative energy loss calculations~\cite{AJMS} for our studies. The radiative energy loss calculations are further discussed in \ref{a1:radloss}.

\begin{figure}[htb!]
	\centering
		\includegraphics[width=0.46\textwidth]{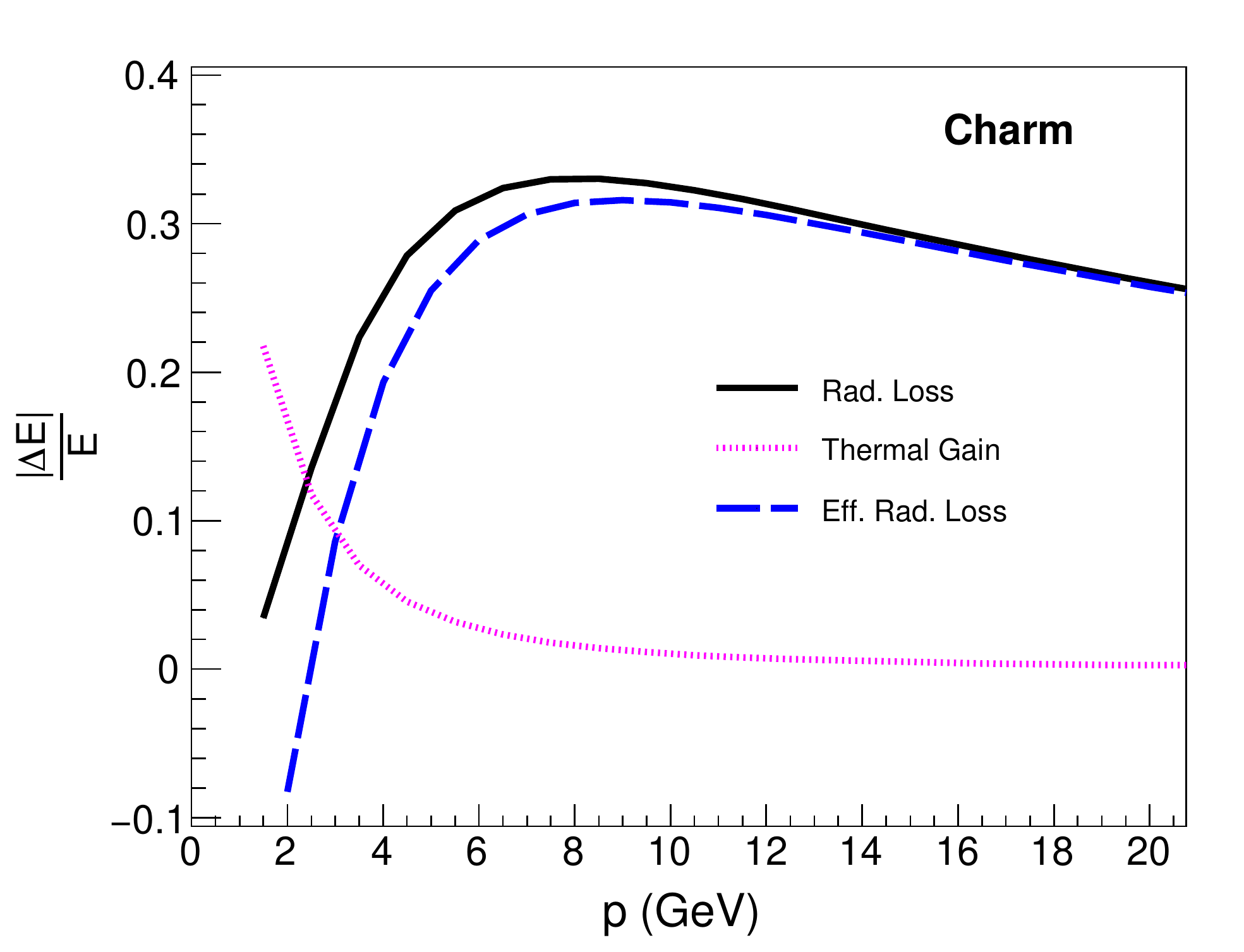}
		\includegraphics[width=0.46\textwidth]{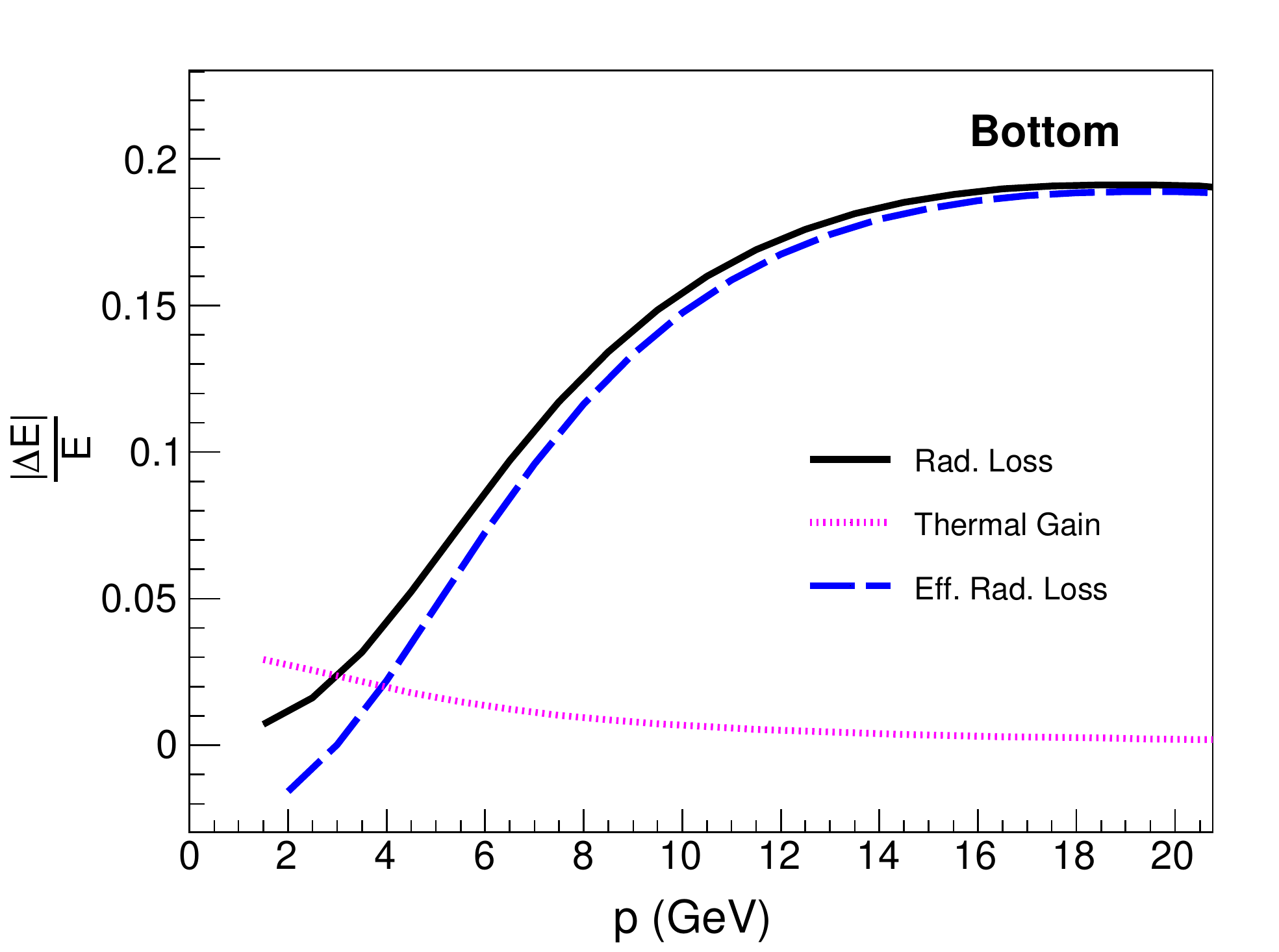}
		\caption{Fractional energy loss and gain of a charm (upper plot) and a bottom (lower plot) quark inside the QGP due to radiative energy loss~\cite{AJMS} and thermal gluon absorption~\cite{wangwang} as a function of momentum $\rm{p}$. The path length considered here is $L=5$ fm.}
		\label{delEbyE_rad_cb}
\end{figure}

Figure~\ref{delEbyE_rad_cb} displays the fractional energy loss in radiative processes~\cite{AJMS}, and also the fractional energy gain due to thermal gluon absorption~\cite{wangwang} for charm (upper plot) and bottom (lower plot) quarks as a function of momentum $p$. It is clear that the energy gain for both charm and bottom quarks due to thermal gluon absorption is relatively more at the lower momentum region (2–10 GeV) than at the higher momentum ($>10$ GeV) region. This means that the thermal gluon absorption and thus the energy gain due to this becomes substantial only in the low momentum limit of the heavy quarks. The effect of the thermal gluon absorption is to reduce the radiative energy loss up to a certain value of momentum, beyond which their contributions gradually diminish. We call the reduced radiative energy loss as the effective radiative energy loss.

\subsubsection{Collisional energy gain}
In this section, we discuss another important mechanism that influences heavy quark propagation in the medium. 
The heavy quarks collide elastically with the particles of the medium during propagation and lose energy. The collisional energy loss per unit length ($-dE/dx$) was calculated in the past by several authors\cite{TG,BT,Alex,PP}. 
The calculations of collisional energy loss consider the QGP medium in an average manner, i.e., without considering microscopic fluctuations. Nevertheless, QGP is a statistical system of mobile charged partons which could be characterised by widespread stochastic field fluctuations. These lead to fluctuations in the velocity of the propagating heavy quarks correlated to the fluctuations in the colour field. Such correlations give rise to collisional energy gain of the heavy quarks as argued in Ref.~\cite{Fl}.
%Since  the energy loss is of topical interest for the
%phenomenology of  heavy quark  jet quenching in hot and dense medium.
 This energy gain is estimated by using semi-classical approximation, and the leading-log (LL) contribution can be expressed as~\cite{Fl},
 \begin{eqnarray} 
 \left(\frac{dE}{dx}\right)_{\mbox{fl}}^{\mbox{LL}} = 2\pi C_F\alpha_{s}^{2}\left(1+\frac{n_f}{6}\right)\frac{T^3}{Ev^2}\ln{\frac{1+v}{1-v}} \ln{\frac{k_{\mbox{max}}}{k_{\mbox{min}}}},
 \end{eqnarray}
 where $k_{\mbox{min}} = \mu_g $ is the Debye mass and
 \begin{eqnarray*} 
 %\nolinenumbers
 k_{\mbox{max}} = \mbox{min}\left[E, {2q(E+p)}/{\sqrt{M^2+2q(E+p)}} \right],
 \end{eqnarray*} with $q \sim T$ is the representative  momentum of the thermal partons. 
 
\begin{figure}[htb!]
	\centering
		\includegraphics[width=0.45\textwidth]{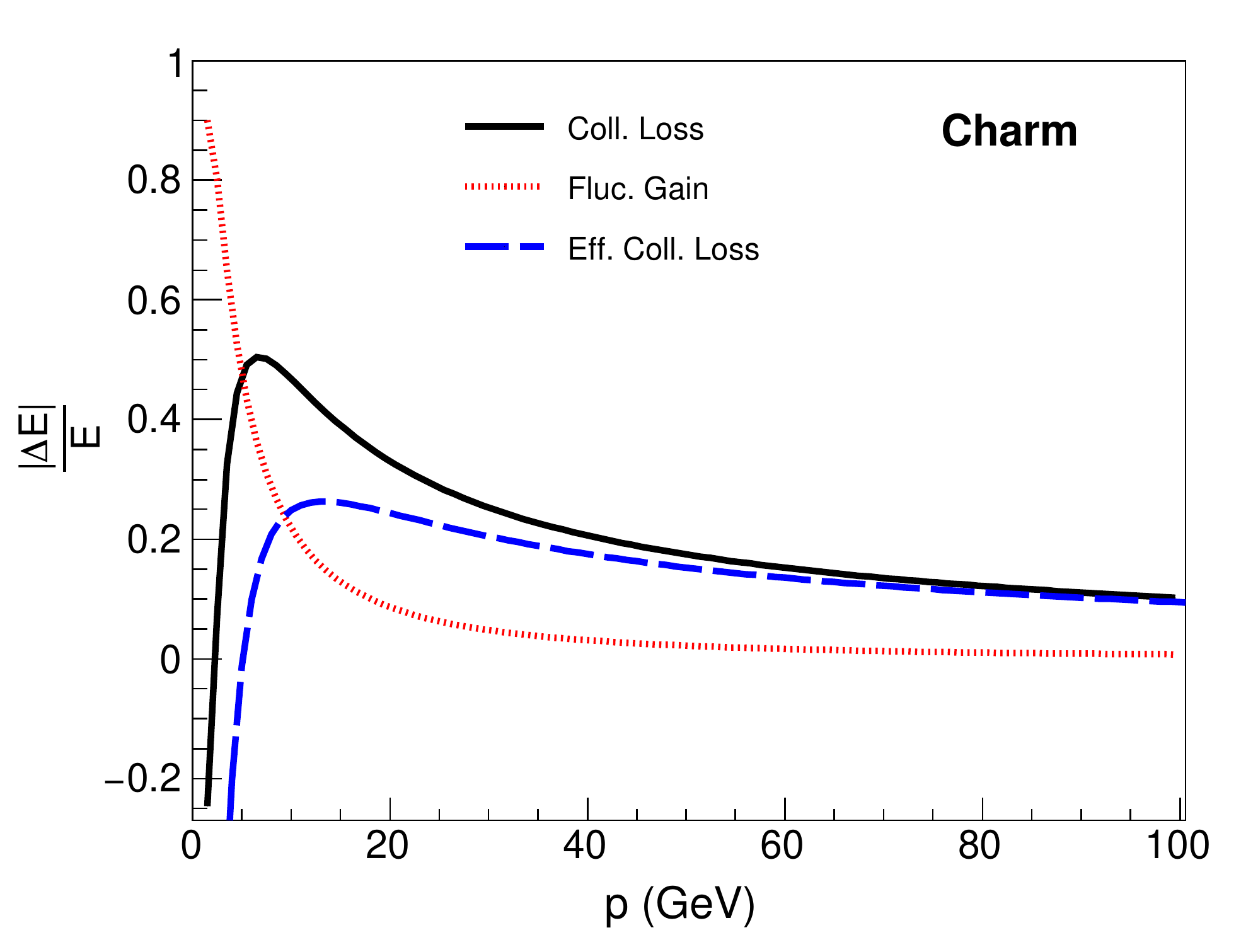}
		\includegraphics[width=0.45\textwidth]{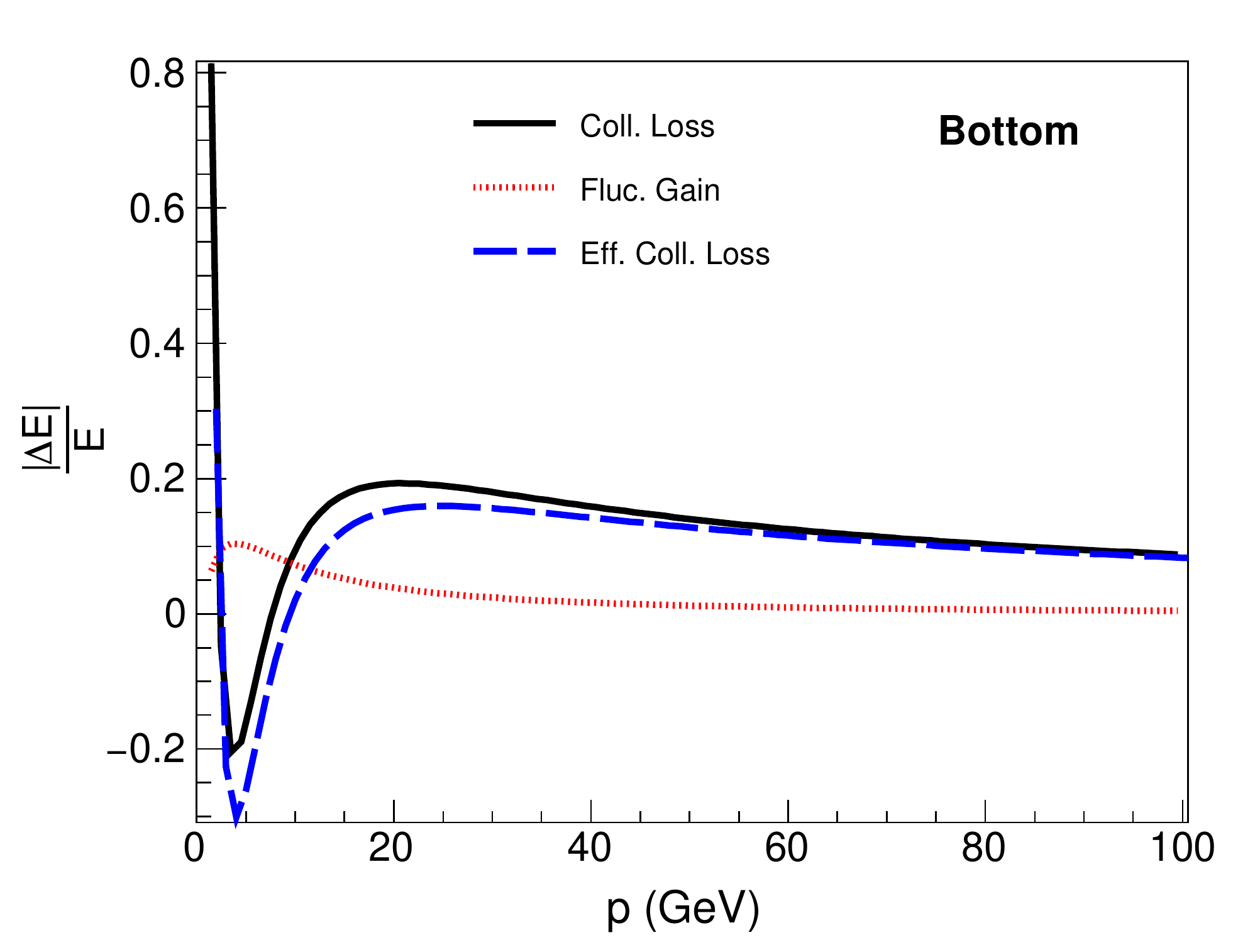}
		\caption{Fractional energy loss and gain of a charm (upper plot) and a bottom (lower plot) quark inside the QGP due to collisional energy loss~\cite{TG} and field fluctuations~\cite{Fl} as a function of momentum $\rm{p}$. The path length considered here is $L=5$ fm.}
		\label{delEbyE_col_cb}
\end{figure}

In the present work, we incorporate the effect of these fluctuations, i.e., we consider the collisional energy gain due to the fluctuations~\cite{Fl} along with the heavy quark collisional energy loss~\cite{TG}. For further discussions on collisional energy loss used here, see \ref{a1:colloss}.

In Fig.~\ref{delEbyE_col_cb}, we depict the fractional energy loss due to collisional processes~\cite{TG} along with the energy gain due to field fluctuations~\cite{Fl} for charm (upper plot) and bottom (lower plot) quark. The energy gain due to fluctuations is relatively higher at the lower momentum region, both for charm and bottom quarks. This energy gain reduces the energy loss due to collisional processes, and we call this reduced collisional loss as the effective collisional loss. The reduction of collisional energy loss is prominent at the lower momentum region.

\begin{figure}[htb!]
	\centering
		\includegraphics[width=0.45\textwidth]{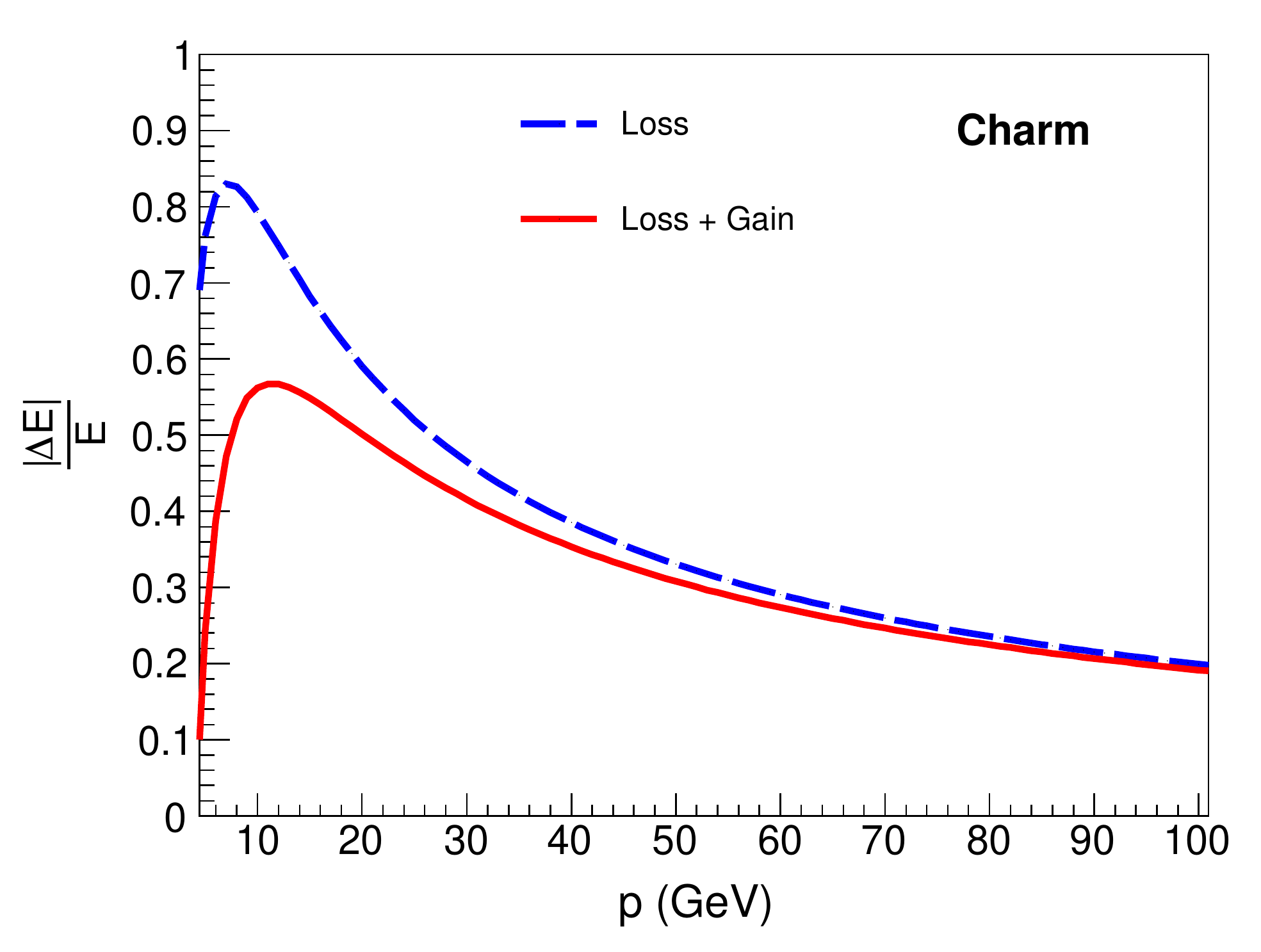}
		\includegraphics[width=0.45\textwidth]{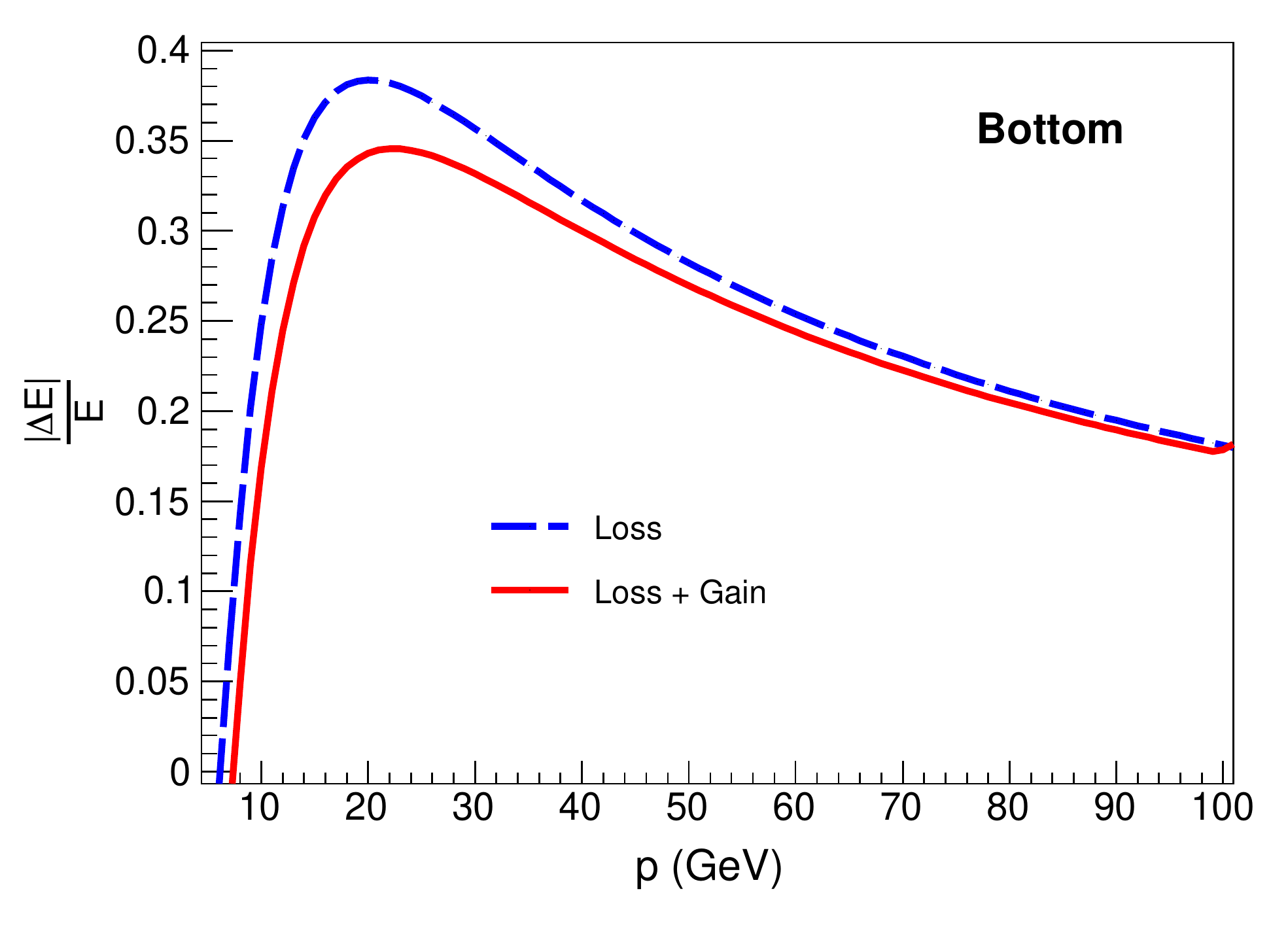}
		\caption{Momentum dependence of total fractional energy loss (radiative+collisional) and loss (radiative+collisional) + gain (thermal+fluctuations) of a charm (upper plot) and a bottom (lower plot) quark inside the QGP. The path length taken here is $L=5$ fm.}
		\label{delEbyE_detbal_cb}
\end{figure}
%\FloatBarrier

Thermal gluon absorption and field fluctuations in the QGP reduce the heavy quark energy loss due to radiative and collisional processes, respectively. Therefore, the total fractional energy loss (radiative+collisional) along with gain (thermal+fluctuations) contribution (red solid curve) is less compared to the case with loss (radiative+collisional) only (blue dashed curve), as shown in Fig.~\ref{delEbyE_detbal_cb}. Since both the energy gains are significant at the lower momentum region, the effect of them is very much prominent for the charm and the bottom quarks of lower momenta.

\section{Results and Discussions}
\label{sec3}
After incorporating different underlying mechanisms of heavy quark energy loss and gain in the medium, we turn to estimate the experimental observables. The nuclear modification factor $R_{AA}$ of heavy flavour mesons,  viz., $D$ and $B$ mesons have been calculated. We employ the fragmentation mechanism in order to obtain heavy flavour mesons from heavy quarks as discussed earlier in Sec.~\ref{sec2.1}.

\begin{figure}[htb!]
	\centering
	\includegraphics[scale=0.4]{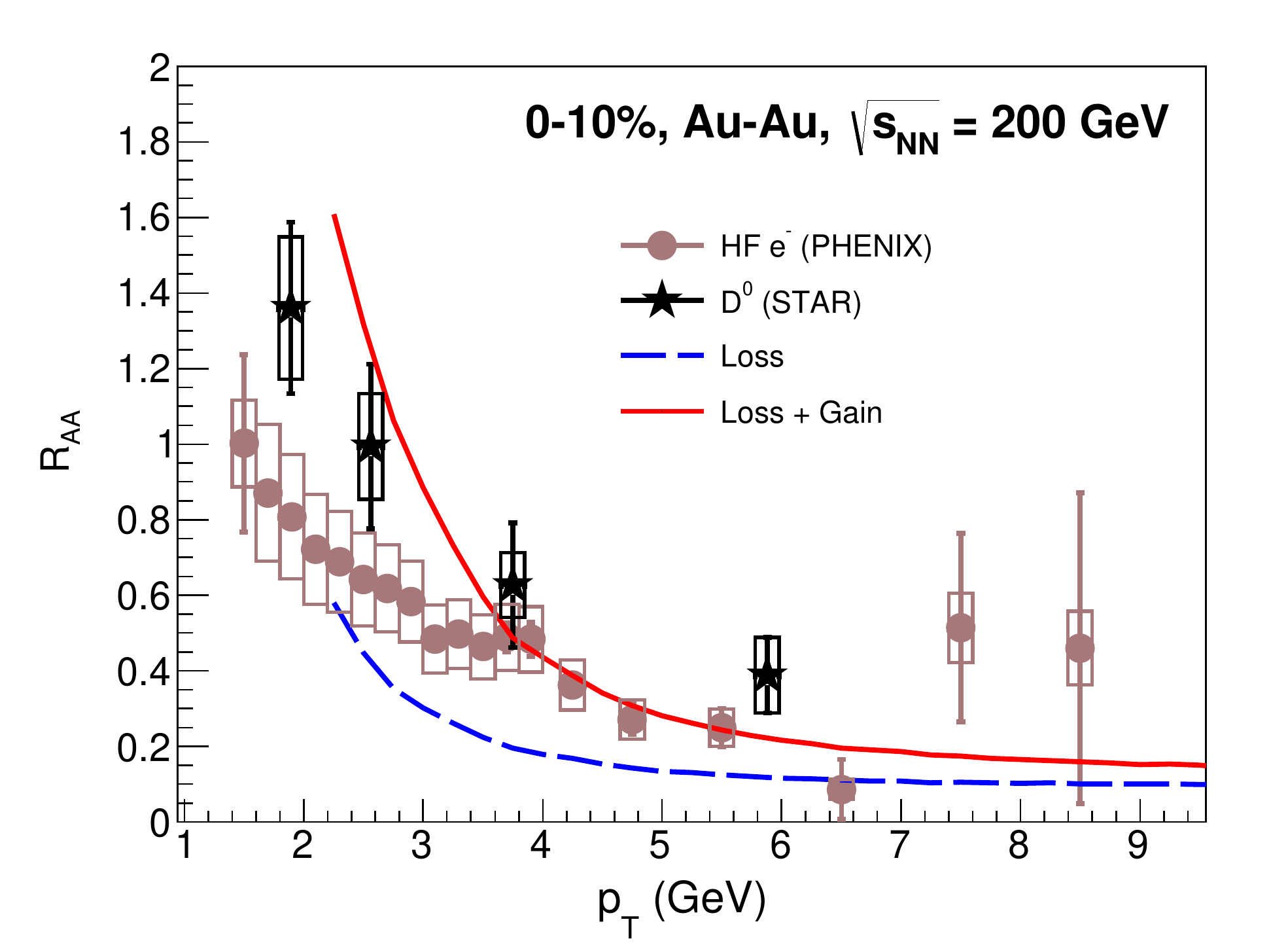}
	\caption{The nuclear modification factor, $R_{AA}$, of $D^0$ mesons with considering the charm quark energy loss (collisional~\cite{TG}+radiative~\cite{AJMS}) and gain (thermal gluon absorption~\cite{wangwang}+field fluctuations~\cite{Fl}), as a function of transverse momentum $p_T$ for 0-10\% centrality in Au-Au collisions at $\sqrt {s_{NN}} = 200$ GeV. The experimental data of $R_{AA}$ for $D^0$ mesons and heavy flavour decay electrons are taken from STAR~\cite{Rhic2} and PHENIX~\cite{Rhic1} experiments, respectively.}
	\label{raa_d_rhic}
\end{figure}

Figure~\ref{raa_d_rhic} depicts the nuclear modification factor, $R_{AA}$, of $D^0$ mesons as a function of $p_T$ for 0-10\% centrality in Au-Au collisions at $\sqrt {s_{NN}} = 200$ GeV, with and without considering the energy gain (thermal+fluctuations) along with the energy loss (radiative+collisional) processes. The calculations are compared to the experimental measurements of $R_{AA}$ for $D^0$ mesons and heavy flavour decay electrons from STAR~\cite{Rhic2} and PHENIX~\cite{Rhic1} experiments, respectively. We observe that there is a significant effect of the energy gain on $D^0$ meson $R_{AA}$ at $\sqrt {s_{NN}} = 200$ GeV in RHIC. It has been argued earlier that the effect of the gain is important at the lower momenta which is also manifested in the calculations as shown in Fig.~\ref{raa_d_rhic}.

%\FloatBarrier

\begin{figure}[htb!]
	\centering
	\includegraphics[scale=0.4]{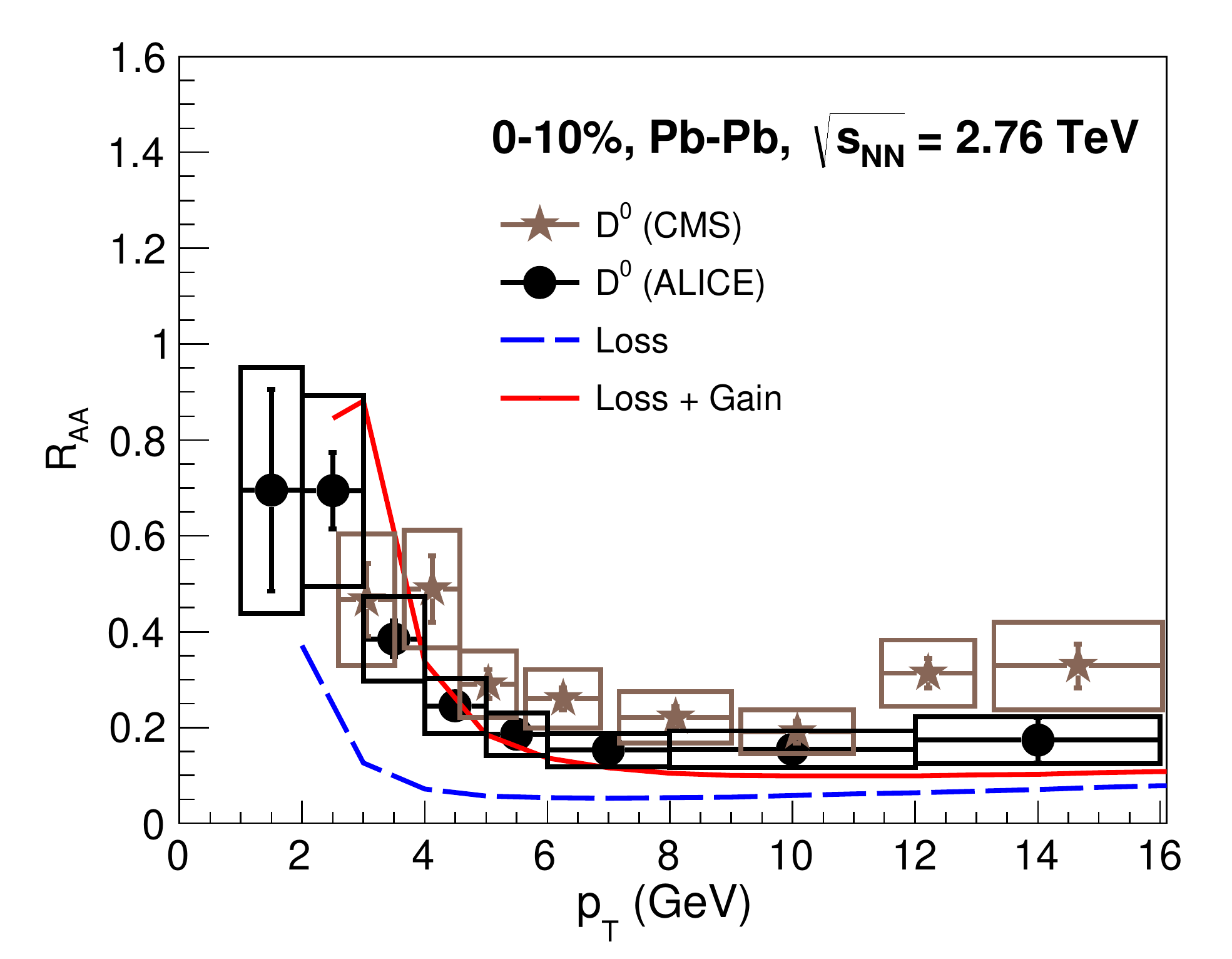}
	\caption{The nuclear modification factor, $R_{AA}$, of $D^0$ mesons with considering the charm quark energy loss (collisional~\cite{TG}+radiative~\cite{AJMS}) and gain (thermal gluon absorption~\cite{wangwang}+field fluctuations~\cite{Fl}), as a function of transverse momentum $p_T$ for 0-10\% centrality in Pb-Pb collisions at $\sqrt {s_{NN}} = 2.76$ TeV. The experimental measurements are taken from ALICE~\cite{ALICE_D} and CMS~\cite{CMS_D_2TeV} experiments.}
	\label{raa_d_10_2tev}
\end{figure}

\begin{figure}[htb!]
	\centering
	\includegraphics[scale=0.4]{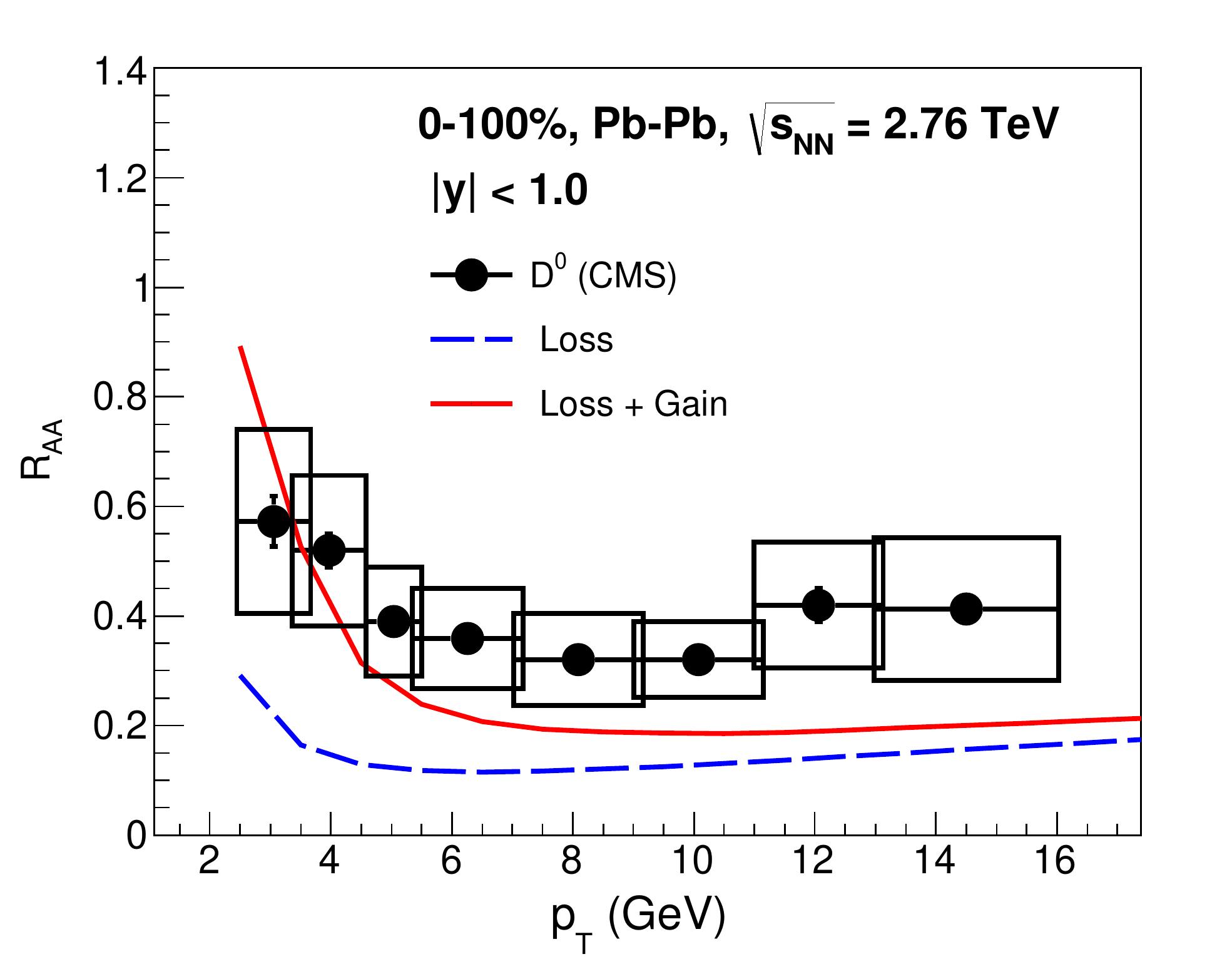}
	\caption{The nuclear modification factor $R_{AA}$ of $D^0$ mesons with energy loss (collisional~\cite{TG}+radiative~\cite{AJMS}) and gain (thermal gluon absorption~\cite{wangwang}+field fluctuations~\cite{Fl}), as a function of transverse momentum $p_T$ for 0-100\% centrality in Pb-Pb collisions at $\sqrt {s_{NN}} = 2.76$ TeV. The experimental measurements are obtained from CMS experiment~\cite{CMS_D_2TeV}.}
	\label{raa_d_2tev}
\end{figure}

Figures~\ref{raa_d_10_2tev} and \ref{raa_d_2tev} display the nuclear modification factor $R_{AA}$ as a function of $p_T$ for $D^0$ mesons in Pb-Pb collisions at $\sqrt {s_{NN}} = 2.76$ TeV for 0–10\% and 0–100\% centralities, respectively, obtained with and without considering the energy gain (thermal+fluctuations). The results are compared with the experimental measurements from ALICE~\cite{ALICE_D} and CMS~\cite{CMS_D_2TeV} experiments. The calculations without the effect of gain show large suppression whereas consideration of the gain produces less suppression. The calculations imply that the gain effect is important for the low $p_T$ $D^0$ mesons.
%The experimental results are described within their uncertainties when the detailed balance of energy loss and gain are taken into consideration.

\begin{figure}[htb!]
	\centering
	\includegraphics[scale=0.4]{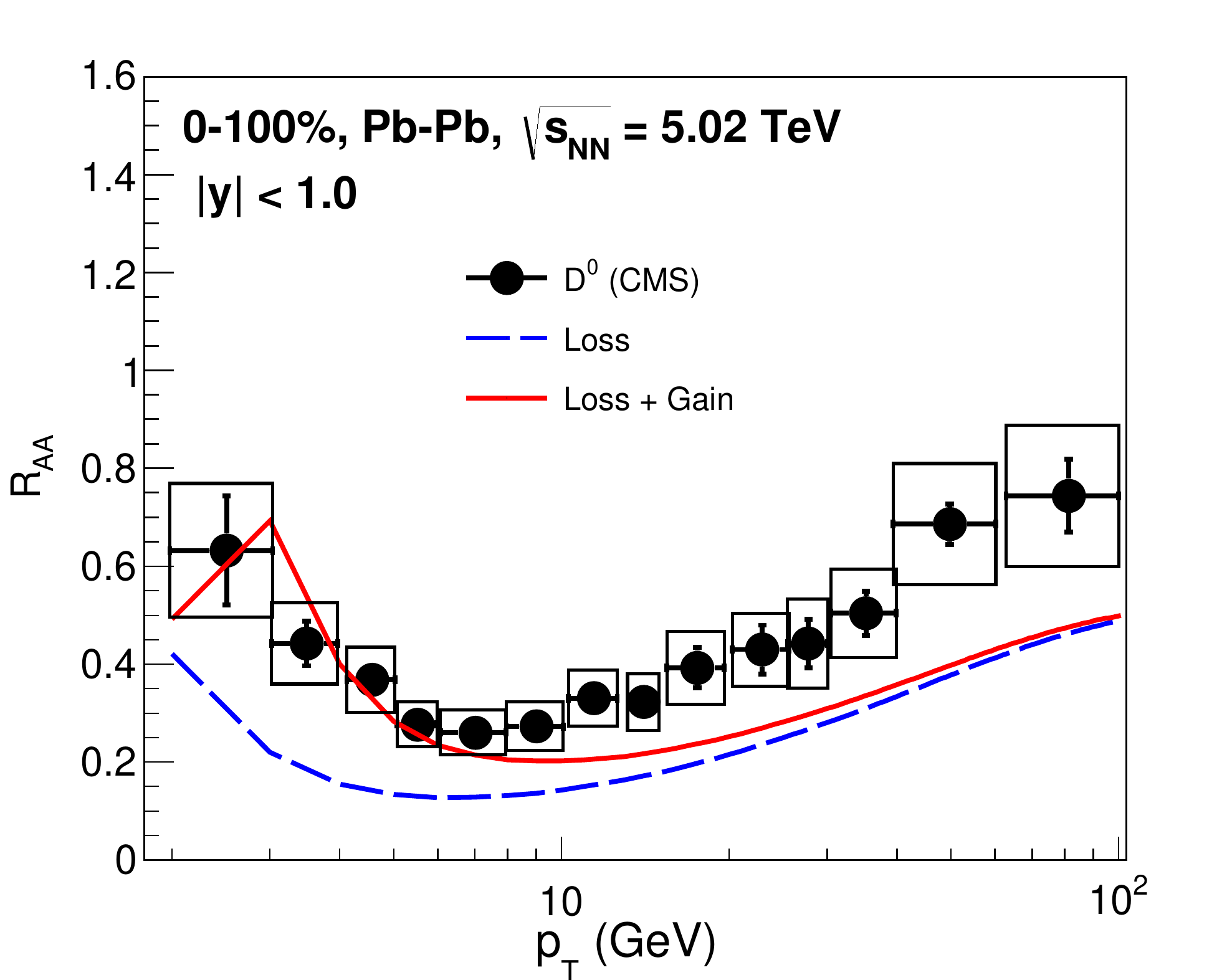}
	\caption{The nuclear modification factor $R_{AA}$ of $D^0$ mesons with taking care of charm quark energy loss (collisional~\cite{TG}+radiative~\cite{AJMS}) and gain (thermal gluon absorption~\cite{wangwang}+field fluctuations~\cite{Fl}), as a function of transverse momentum $p_T$ for 0-100\% centrality in Pb-Pb collisions at $\sqrt {s_{NN}} = 5.02$ TeV. The experimental measurements are obtained from CMS experiment~\cite{CMS_D_5TeV}.}
	\label{raa_d_5tev}
\end{figure}

\begin{figure}[htb!]
	\centering
	\includegraphics[scale=0.4]{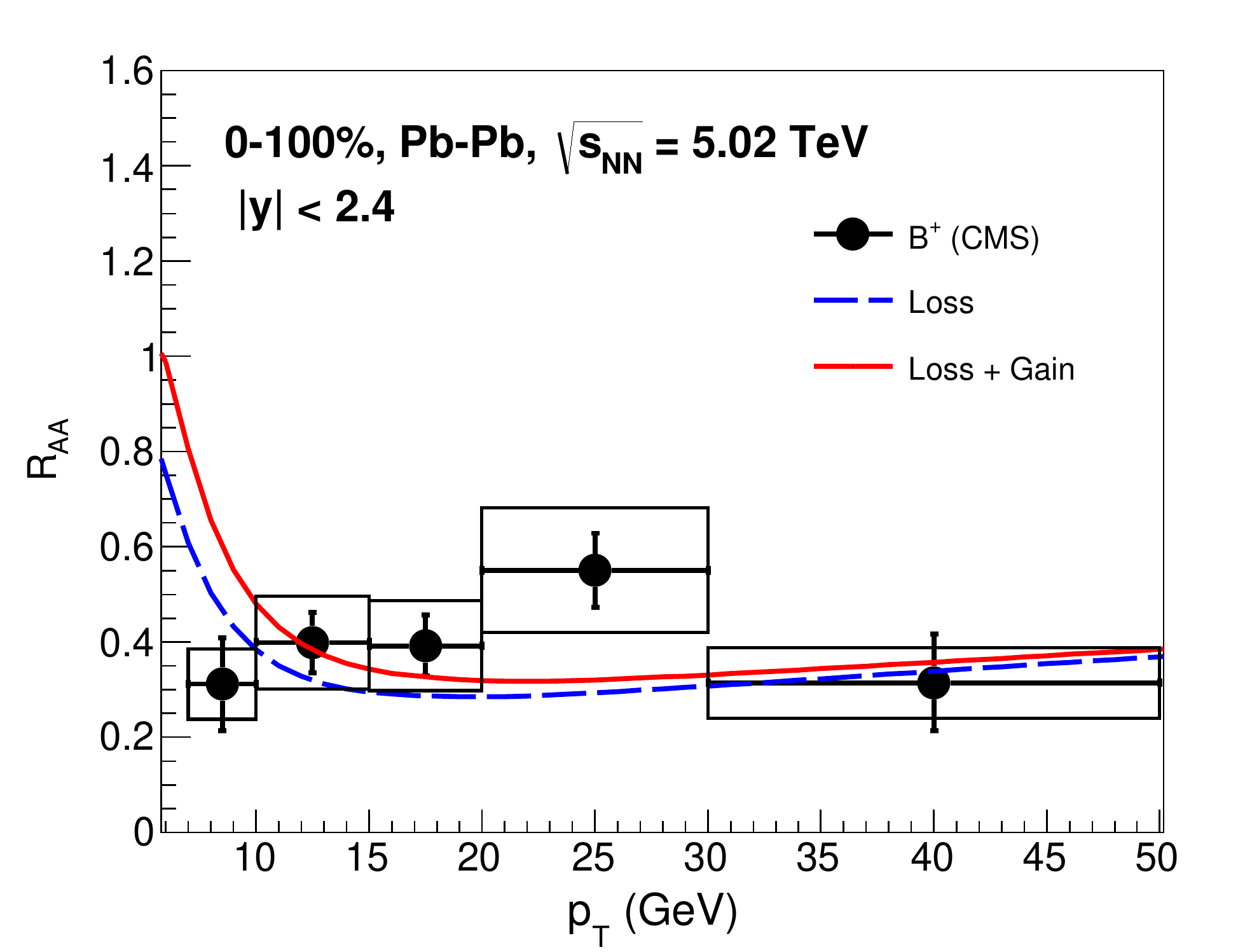}
	\caption{The nuclear modification factor, $R_{AA}$, of $B^+$ mesons with considering the bottom quark energy loss (collisional~\cite{TG}+radiative~\cite{AJMS}) and gain (thermal gluon absorption~\cite{wangwang}+field fluctuations~\cite{Fl}) as a function of transverse momentum $p_T$ for 0-100\% centrality in Pb-Pb collisions at $\sqrt {s_{NN}} = 5.02$ TeV. The experimental measurements are obtained from CMS experiment~\cite{CMS_B_5TeV}.}
	\label{raa_b}
\end{figure}

In Figs~\ref{raa_d_5tev} and ~\ref{raa_b}, the $p_T$ dependent nuclear modification factor, $R_{AA}$, of $  D^0$ and $B^+$ mesons in Pb-Pb collisions at $\sqrt {s_{NN}} = 5.02$ TeV for 0–100\% centrality have been shown, respectively. The calculations of $R_{AA}$ are made with and without considering the energy gain (thermal+fluctuations) by charm and bottom quarks. Here, we compare the calculations with the experimental data of $R_{AA}$ for $D^0$~\cite{CMS_D_5TeV} and $B^+$~\cite{CMS_B_5TeV} mesons from CMS experiment. At the low $p_T$ region, the calculations are close to the experimental measurement of $R_{AA}$ for $D^0$ mesons, when we consider the charm quarks lose and gain energy in the medium. However, it is noteworthy to mention here that the calculations are sensitive on different model parameters used here. The calculations are away form the measured $R_{AA}$ at the high $p_T$ region. Improved energy loss calculations in this region may shed light on this. Results for $B^+$ mesons, with and without considering the energy gain, are very close to experimental measurements within the uncertainties, though the experimental uncertainties are quite large. It is observed that the effect of the energy gain is also prominent at low momentum for $B^+$ mesons and the effect is less compared to the suppression of $D^0$ mesons.

We emphasize here that the thermal absorption and the field fluctuations are found to play a key role in the heavy flavour suppression at the top RHIC and the LHC energies. The approximations and possible sources of uncertainties in our calculations are listed below: 

\begin{enumerate}
\item A simplistic fireball model has been applied for the QGP medium evolution. There are uncertainties in the initial conditions for the medium evolution in this model. The thermalization of the QGP is not yet fully understood. The initial temperature $T_0$ and the effective temperature of the medium may vary, since initial time $\tau_0$ may vary between 0.1 and 0.5 fm/c.

\item The axial gauge and the leading-log approximation have been made in the estimation of energy gain due to thermal gluon absorption.  

\item The mean energy loss and gain (caused by field fluctuations) calculations have been done by using semi-classical approximation, which is equivalent to the Hard Thermal Loop approximation based on the weak coupling limit\cite{TG,BT,PP,AJMS}.

\item A constant momentum and temperature independent coupling constant is used instead of using running coupling.

\item We used the fragmentation mechanism without recombination in the heavy flavour hadronization process. There are uncertainties in parton distribution and fragmentation function which eventually might affect the calculations of $R_{AA}$.

%\item Coalescence scenario along with fragmentation in heavy flavour hadronization is not considered. This could influence the $R_{AA}$.

\end{enumerate}

\section{Conclusion}
\label{sec4}
Parton energy loss in a QGP medium leads to high $p_T$ hadron suppression or jet quenching, and an observation of such a phenomenon uncovers many dynamical properties of that medium. The QGP, being a hot and statistical system of coloured partons, thermally generated gluons and colour field fluctuations are present in the medium. In the presence of such thermal gluons and field fluctuations in the medium, a heavy quark gains and loses energy while passing through the medium. The effect of thermal gluon absorption by a heavy quark in the QGP where the generated coloured field is fluctuating in nature is crucial for phenomenological study of heavy flavour suppression. Here, we have taken care of the energy gain due to thermal absorption and field fluctuations along with the energy loss caused by the collisions and gluon radiations by a propagating heavy quark. The nuclear modification factor $R_{AA}$ of $D$ mesons in Au-Au collisions at $\sqrt{s_{NN}} = 200$ GeV, in Pb-Pb collisions at $\sqrt{s_{NN}} = 2.76$ and $5.02$ TeV, and that of $B$ mesons at $\sqrt{s_{NN}} = 5.02$ TeV have been computed by including the energy gain and loss processes in the medium. We find that the calculations for $D$ and $B$ mesons with the effect of thermal gluon absorption and field fluctuations along with the usual loss processes (collisional+radiative) produce less suppression compared to the case where only usual loss processes are considered. Nevertheless, the calculations for $D$ mesons in Pb-Pb collisions at $\sqrt{s_{NN}} = 5.02$ TeV show larger suppression compared to experimental data at the high $p_T$ region. The effect of the thermal absorption and medium fluctuations is less for $B$ mesons compared to $D$ mesons. Results for $B$ mesons with and without considering the energy gains are close to the data within large uncertainties. Although the model calculations here are sensitive to different parameters. Future precise measurements and better constrained model calculations would help us understand the different energy gain processes and their consequences on heavy flavour observables. The presence of thermal gluons and medium fluctuations in the hot QGP is found to play a pivotal role in driving the heavy flavour suppression at the lower momenta in the RHIC and LHC energies.

\section*{Acknowledgements}
I acknowledge support from the Office of Nuclear Physics within the US DOE Office of Science, under Grant DE-FG02-89ER40531. I am thankful to Declan Keane, Prithwish Tribedy, Zubayer Ahammed and Md Hasanujjaman for carefully reading the manuscript and the suggestions.

%\pagebreak

\appendix
\section{Radiative energy loss: Generalized dead cone approach}
\label{a1:radloss}
The expression for heavy quark radiative
energy loss, based on the generalised dead cone approach and the gluon emission probability~\cite{abir12}, can be found as~\cite{AJMS}:
\begin{equation}
%\begin{split}
\frac{dE}{dx} = 24\alpha_{s}^{3}\rho_{QGP}\frac{1}{\mu_g}\left(1-\beta_1\right) \left(\sqrt{\frac{1}{(1-\beta_1)}\log(\frac{1}{\beta_1})}-1\right) \mathcal F(\delta),
%\end{split}
\end{equation}
 with
\begin{eqnarray}
\mathcal F(\delta) = 2\delta - \frac{1}{2}\log\left(\frac{1+\frac{M^2}{s}e^{2\delta}}{1+\frac{M^2}{s}e^{-2\delta}}\right)- \left(\frac{\frac{M^2}{s}\sinh{(2\delta)}}
{1+2\frac{M^2}{s}\cosh{(2\delta)}+\frac{M^4}{s^2}} \right),
\end{eqnarray}
where
\begin{equation}
\delta = \frac{1}{2}\log\left[\frac{1}{(1-\beta_1)}\log\left (\frac{1}{\beta_1}\right )\left(1+\sqrt{1-\frac{(1-\beta_1)}{\log\frac{1}{\beta_1}}}\right)^2\right]
\end{equation}
and $\rho_{QGP}$  is the density of the QGP medium which acts as a background
containing the target partons. If $\rho_q$ and $\rho_g$ are the density
of quarks and gluons respectively in the medium, then the $\rho_{QGP}$ is given by
\begin{equation}
\rho_{QGP} = \rho_q+\frac{9}{4}\rho_g,
\end{equation}
\begin{equation}
\beta_1 = \frac{\mu_{g}^{2}}{CET},
\end{equation}
\begin{equation}
 C = \frac{3}{2}-\frac{M^2}{4ET}+\frac{M^4}{48E^2T^2\beta_0}\log\left(\frac{M^2+6ET(1+\beta_0)}{M^2+6ET(1-\beta_0)}\right),
\end{equation}
\begin{equation} 
\beta_0 = \sqrt{1-\frac{M^2}{E^2}}
\end{equation}

It is to be noted here that later a kinematical correction of the above calculations was made in Ref.\cite{KPV}.

\section{Collisional energy loss: Thoma Gyulassy approach}
\label{a1:colloss}

The collisional energy loss per unit length $dE/dx$ of a heavy quark of mass $m$, momentum $p$, energy $E=\sqrt{p^2+m^2}$ and velocity $v=p/E$ has been reported by Thoma and Gyulassy~\cite{TG}:

\begin{eqnarray}
\frac{dE}{dx} = \frac{4\pi\alpha_{s}^{2}T^{2}C_F}{3} \ln\left(\frac{k_{max}}{k_D}\right)\frac{1}{v^2}\left(v+\frac{v^2-1}{2}\ln\frac{1+v}{1-v}\right)
\end{eqnarray}
where $k_{max} \approx \frac{4Tp}{E-p+4T}$ is maximal momentum transfer, with $T$ being the temperature of the medium and $k_D = \sqrt{3}m_g$ is Debye momentum, $m_g$ being thermal gluon mass.

\end{document}